# Collaborative Learning with Artificial Intelligence Speakers (CLAIS): Pre-Service Elementary Science Teachers' Responses to the Prototype

Gyeong-Geon Lee[a,b,c,d]*, Seonyeong Mun[c]*, Myeong-Kyeong Shin[d] and Xiaoming Zhai[a,b]

[a]*AI4STEM Education Center, University of Georgia, Athens, GA 30602, United States*
[b]*Department of Mathematics, Science, and Social Sciences Education, University of Georgia, Athens, GA 30602, United States*
[c]*Department of Chemistry Education, Seoul National University, 1 Gwanakgu Gwanakro, Seoul 08826, Republic of Korea;*
[d]*Department of Science Education, Gyeongin National University of Education, 62 Gyeyanggu Gyesanro, Incheon 21044, Republic of Korea*

*corresponding author: crusaderlee@snu.ac.kr; suyo2478@naver.com

## Abstract

This research aims to demonstrate that AI can function not only as a tool for learning, but also as an intelligent agent with which humans can engage in collaborative learning (CL) to change epistemic practices in science classrooms. We adopted a design and development research approach, following the Analysis, Design, Development, Implementation and Evaluation (ADDIE) model, to prototype a tangible instructional system called Collaborative Learning with AI Speakers (CLAIS). The CLAIS system is designed to have 3–4 human learners join an AI speaker to form a small group, where humans and AI are considered as peers participating in the Jigsaw learning process. The development was carried out using the NUGU AI speaker platform. The CLAIS system was successfully implemented in a *Science Education* course session with 15 pre-service elementary science teachers. The participants evaluated the CLAIS system through mixed methods surveys as teachers, learners, peers, and users. Quantitative data showed that the participants' Intelligent-Technological, Pedagogical, And Content Knowledge was significantly increased after the CLAIS session, the perception of the CLAIS learning experience was positive, the peer assessment on AI speakers and human peers was different, and the user experience was ambivalent. Qualitative data showed that the participants anticipated future changes in the epistemic process in science classrooms, while acknowledging technical issues such as speech recognition performance and response latency. This study highlights the potential of Human-AI Collaboration for knowledge co-construction in authentic classroom settings and exemplify how AI could shape the future landscape of epistemic practices in the classroom.

**Acknowledgement**

This research is supported by Seoul National University Department of AI-Integrated Education

**Statements and Declarations**

The authors have no relevant financial or non-financial interests to disclose.

# Collaborative Learning with Artificial Intelligence Speakers (CLAIS) : Pre-Service Elementary Science Teachers' Responses to the Prototype

**Abstract**

This research aims to demonstrate that AI can function not only as a tool for learning, but also as an intelligent agent with which humans can engage in collaborative learning (CL) to change epistemic practices in science classrooms. We adopted a design and development research approach, following the Analysis, Design, Development, Implementation and Evaluation model, to prototype a tangible instructional system called Collaborative Learning with AI Speakers (CLAIS). The CLAIS system is designed to have 3–4 human learners join an AI speaker to form a small group, where humans and AI are considered as peers participating in the Jigsaw learning process. The development was carried out using the NUGU AI speaker platform. The CLAIS system was successfully implemented in a *Science Education* course session with 15 pre-service elementary science teachers. The participants evaluated the CLAIS system through mixed methods surveys as teachers, learners, peers, and users. Quantitative data showed that the participants' Intelligent-Technological, Pedagogical, And Content Knowledge was significantly increased after the CLAIS session, the perception of the CLAIS learning experience was positive, the peer assessment on AI speakers and human peers was different, and the user experience was ambivalent. Qualitative data showed that the participants came to anticipate future changes in the epistemic process in science classrooms, while acknowledging technical issues such as speech recognition and response latency. This study highlights the potential of Human-AI Collaboration for knowledge co-construction in authentic classroom settings and exemplify how AI could shape the future landscape of epistemic practices in the classroom.

**1. Introduction**

The 21st century has seen a rapidly changing world with global trends of digitalisation and advances in Artificial Intelligence (AI). While the concept of AI can be traced back to the early 20th century, the development of machine learning (ML) algorithms based on deep neural networks, leveraging hardware capacity, and the accumulation of data, has led to the widespread use of AI recently (Norvig & Russell, 2021). As of the 2020s, AI technologies are not only transforming our living world but also urging educational scholars to rethink both the goals and methods of education to prepare learners who will live in the future society (OECD, 2019; Hwang et al., 2020; Homes et al., 2019). In line with this, the *call for papers* in the *Science & Education* journal[1] suggests we need to examine how the Digital Age and AI will change epistemic beliefs and practices in science education.

There is the framework in the above discourse that AI as a new driving force will transform pedagogical conventions. Therefore, to respond to this *call for papers*, it is necessary to reflect on the current epistemic beliefs and practices in science education and what kind of AI technology could be integrated into science classrooms as a facilitator of educational change. In other words, it is recommended to explore the intersection of established teaching and learning practices in science education and novel advances in AI, so that we can interpret the evolving educational phenomena using a familiar language.

Within this framework, this design and development (D&D) case study of an instructional system explored a locus where collaborative learning (CL), a well-known teaching and learning method in science education, intersects with AI in Education (AIEd), a relatively new research area. In the following sections, we reviewed the two

---

[1] https://www.springer.com/journal/11191/updates/23312040 (Retrieved January 5th, 2023)

strands of research –CL and AIEd and then introduced the Human-AI collaboration (HAC) perspective. Based on these theoretical underpinnings, we proposed a palpable instructional system called Collaborative Learning with AI Speaker (CLAIS). By exploring the impact of AI on CL in the context of science teacher education, this research provides insight into the potential of HAC for the future of science education.

*1.1 Constructivist Theory and Collaborative Learning*

Constructivist learning theories have emphasised the interaction between learners and the environment surrounding them as the essential process of learning. In a Piagetian sense, the interaction facilitates assimilation and accommodation in the individual learner's mind. According to Vygotsky's (1978) sociocultural theory, peer interaction is even more emphasised because it scaffolds students' learning in the zone of proximal development (ZPD) (Blosser, 1993; Bowen, 2000). Therefore, CL, a form of peer learning (Damon & Pehlps, 1989) defined as "a *situation* in which *two or more* people *learn* or attempt to learn something *together*" (Dillenbourg, 1999, italics in original), has been one of the most important learning methods in science education from kindergarten to university.

As expected, many theoretical and empirical studies have focused on the ways and consequences of structuring and formulating learners' interactions in CL. There has been a discussion about the name of the activity system, and many scholars have used 'collaboration' and 'cooperation' interchangeably (Htilz, 1998; Johnson & Johnson, 2002; O'Donnell & Hmelo-Silver, 2013). Some argue that 'collaboration' focuses more on the reciprocity of influence and the learning processes through linguistic interactions between heterogeneous learners, and 'cooperation' puts more emphasis on individual responsibility and learning outcomes of homogenous learners (Damon & Phelps, 1989; Dillenbourg, 1999; O'Donnell & Hmelo-Silver, 2013; Authors, 2021). Therefore, 'collaboration' is a better fit for this study, aligning with the goal of investigating the reciprocity relationships between AI and students, as well as the potential influence of AI on students' epistemic process of knowledge building.

The sociocultural perspective further suggests that mutual interactions of heterogeneous group members can contribute to knowledge creation while providing multiple ZPDs, where students "capitalise on their complementary knowledge and expertise and jointly achieve higher level collaborative objectives" (Hakkarainen et al., 2013, p. 66). Hence, studies have investigated the effects of different grouping methods in CL (e.g., heterogeneous, homogeneous, female-dominated, male-dominated) on student performance, with heterogeneous grouping being the accepted recommendation (O'Donnell & Hmelo-Silver, 2013).

Meanwhile, specific CL models such as Jigsaw, Student Team Achievement Division, Group Investigation, and Learning Together have been developed and tested in different classroom conditions (O'Donnell & Hmelo-Silver, 2013). Meta-analyses have shown that these CL methods have a positive impact on students' cognitive and affective learning (Bowen, 2000; Kyndt et al., 2013), which is rationalised by their theoretical underpinnings. For example, in the jigsaw model (Aronson et al., 1978; Slavin, 1978), up to six students form a home group. The content of the lesson is divided into segments and each student is assigned to study one segment. An expert group for each task is temporarily formed by including one student from each home group and providing them with materials and time to study. Students then return to their home groups to share what they have learnt from the expert groups, completing each home group's comprehensive learning of the content. Because of its highly structured characteristics, Jigsaw is considered appropriate for cognitive learning at the university level as well as at the K-12 level (Millis & Cottell, 1998) and for computer-mediated CL (Schaeffer & Cates, 1996; Weidman & Bishop, 2009).

The recent development of the research field of Computer-Supported Collaborative Learning (CSCL) makes it possible to imagine the use of AI in the context of CL. CSCL originally "emerged from an interest in taking advantage of artificial intelligence in education and in educational research" (Stahl, 2015, p. 340), with a particular emphasis on "structures of interaction at the dialogical, small-group and community level, where language is primarily learned and practiced" (p. 337). Therefore, qualitative investigation of the characteristics of learners' interaction is the main method in CSCL studies. For example, Krange and Ludvigsen (2007) designed and implemented a CSCL learning environment in which four students and a teacher simultaneously studied the scientific concept of protein using a 3D model accessed from individual computers. They analysed how computer tools mediated students' interactions from a Vygotskian perspective and reported that students and a teacher tended to gain knowledge about the procedural aspects of the target domain in the CSCL environment.

*1.2 Development of AIEd and Science Education*

The introduction of computers in education has a long history (Homes et al., 2019). Computer-Assisted Instruction (CAI) of the 1980s had already been reported to be effective in science education (Kinzie et al., 1988), although its visionary goals were unrealistic at the time (Okolo et al., 1993). However, the dramatic development of ML technologies in the 2010s has led to the emergence of AIEd programmes, followed by much theoretical discussion. AIEd brings together AI and learning sciences "to promote the development of adaptive learning environments" and other "tools that are flexible, inclusive, personalised, engaging, and effective" for learning (Luckin et al., 2016, p. 18). To achieve this goal, the AIEd community relies on algorithms that incorporate pedagogical, domain, and learner models to store and analyse learner-generated data (Luckin et al., 2016). Holmes et al. (2019) suggested that AIEd can foster students' ability to solve ill-defined problems by providing individualised feedback based on formative assessment. Hwang et al. (2020) further proposed a contemporary framework for the role of AIEd as an intelligent tutor, intelligent tutee, intelligent learning tool/partner, or policy advisor. Srinivasan (2022) extended even Hwang et al.'s (2020) points to include dynamic course authoring, continuous knowledge acquisition, optimising learning content, optimising cognitive load, and creating and managing learning communities.

In science education, AI has primarily been used for automated assessment of student-written text data (Shin & Shim, 2021), probably because science educators often use open-ended (or constructed) items to reveal students' explanations about the phenomena (Authors, 2020). Pioneering studies have focused on demonstrating that ML models could assess students' responses to science items while their data were being collected in large-scale classroom assessment practices or detached from teaching and learning. For example, automated assessments have been attempted for student concepts of natural selection (Nehm & Ha, 2011), climate change (Zhu et al., 2017; Zhu et al., 2020), and acid-base reactions (Haudek et al., 2012). In addition, a recent study has further explored the possibility of automated assessment in the processing of student-generated hand drawings as well as written responses regarding the particulate nature of matter (Authors, 2023a).

Notably, the recent research on automated assessment in science education has moved beyond examining conceptual understanding gear toward students' uses of knowledge in scientific practices, providing them with individualised feedback to support their learning based on their status with appropriate instructional interventions (Authors, 2019; Zhu et al., 2020; Authors, 2020; Authors, 2023a). Although there remain concerns, especially regarding its ability to be socio-culturally and linguistically sensitive (Li et al., 2023), "AI and formative assessment" has already become a widespread practice, as if "the train has left the station" (Authors, 2023c). This and other research show that the integration of AI in classroom assessment practices is making a growing impact on science teaching and learning.

Especially since the release of ChatGPT, research has reported increasing opportunities to involve AI in science learning beyond assessment practices. For example (Author, 2023) shows how AI and ChatGPT can be used to provide learning guidance and adapt to student's individual needs, including students with learning abilities. Reviews also suggest that AI chatbots can be used to provide opportunities for learners to interact with AI to gain knowledge. This existing evidence suggests that AI is playing an increasingly important role in transforming the teaching and learning paradigm beyond assessment. Given AI's prevalent involvement in education, it is thus critical to examine the extent to which the integration of AI in science classrooms may change teachers' and students' epistemic beliefs and their perceptions about the process of science education.

### *1.3 Human-AI Collaboration and Education*

Human-computer interaction (HCI) has been a major area of research for decades (Berg, 2000). While other general HCI-related essentials such as human factors, usability and interface design have been thoroughly studied, educational HCI studies have emphasised the computer as a medium (Berg, 2000). Meanwhile, Card, Moran & Newell (1983) established the psychology of HCI, understanding the human mind as an information-processing system.

Since the rise of AI technology, the literature has shifted the focus toward Human-AI Interaction (HAI). Lai et al. (2021) reviewed more than 80 empirical studies on Human-AI decision-making in various research fields, including educational studies and found that the number of publications on Human-AI interaction and Human-AI decision-making increased geometrically after the 2010s - the number of papers per 2 years was less than 100 until 2016, but reached more than 1,000 for each research topic in 2020. Decision tasks such as student performance prediction, student admission prediction, student dropout prediction, and law school admission test question answering are especially popular.

Despite the rise of HAI research, the field is concerned about learners' active role when engaging in HAI-based learning. For example, learners are supposed to be active knowledge constructors and gainers, but HAI has been reported to fall short of explainability. Consequently, learners may not be able to fully appreciate the outcomes of a black box. Therefore, researchers have focused on the 'explainability' of AI systems, aiming to support users to understand why an agent exhibits a certain behaviour or decision (Hemmer et al., 2021). An explainable AI (XAI) system provides full potential for the "collaboration between humans and AI" (p. 1). They conceptualise 'hybrid intelligence', where human-AI complementarity is revealed in complementary team performance. Meanwhile, the National Academies of Sciences, Engineering, and Medicine (NASEM, 2022) discussed human-AI teaming in military operations. They also suggested that "research is needed to determine improved methods for supporting collaboration between humans and AI systems" (p. 3).

Researchers have put effort to incorporate HAC in educational contexts to promote student-centred learning. For example, Kim et al. (2022) developed a model of student-AI collaboration, derived from the perspective of leading teachers for AIEd. Their model consists of four components - (1) curriculum, (2) student-AI interaction, (3) environment, and (4) evolution over time - and corresponding 11 themes. Meanwhile, Kim (2022) proposed AI-integrated science education by facilitating epistemic discourse in the classroom. She envisioned that science teachers could introduce AI to students in the science classroom and invite students to engage in epistemic discourse, such as "what are the desired epistemic processes?" (epistemic aims), "what are the criteria used to evaluate whether the aims are achieved?" (epistemic ideals), and "what are the processes to achieve the aims?" (reliable epistemic processes). These studies were theoretically sound but called for empirical studies. Moreover, there has been limited research delving into the CL between humans and AI in authentic teaching and learning situations. It is not clear how students, teachers, and AI played a role in such interactive and collaborative learning environments (Shi & Choi, in press). More research is needed to dig into the ways how users interact with AI and how they perceive the usefulness of HAI, based on their epistemic roles.

The CLAIS system in this study is an attempt to further the HAC initiative in terms of CL and AIEd for science education. We suggest that there are several theoretical underpinnings that support the convergence of CL and AIEd. (1) Cultural-Historical Activity Theory (CHAT), which extends Vygotsky's sociocultural constructivism (Engeström, 1987) and enables 'learning with computers' as a 'division of labour' (Tan, 2019); (2) CSCL theory, which puts the focus on AI (Stahl, 2015); and (3) Actor-Network Theory (ANT), which focuses on the networks and relationships between human and non-human actors that influence the construction of new knowledge (Latour, 2007; Roth & McGinn, 1997).

Therefore, in this study, we developed CLAIS system to study the epistemic roles of users and their perceptions of the AI system in supporting science learning. Specifically, we leverage the possible strength of heterogeneous grouping through the inclusion of AI as a CL group member to generate ZPD where scaffolding occurs for human learners. We used the Jigsaw method to engage pre-service science teachers in an AI-integrated learning system by structuring their interactions. Also, the CLAIS system proposed in this study promotes the notion of a collaborative environment that is aligned with the computer - it is 'AI-participated' CL rather than 'computer-assisted' CL.

*1.4 AI Speakers in Education*

AI speakers are one of the most commercialised AI services in our daily lives, such as Apple Siri, Microsoft Coratna, Amazon Alexa, Google Home and Samsung Bixby that millions of users experience on their devices every day (Șerban & Todericiu, 2020; Terzopoulos & Satratzemi, 2020). Accordingly, the notion of AI Speakers in Education (AISEd) has emerged, which assumes smart speakers as intelligent agents that answer students' questions in natural spoken language (Lovato et al., 2019; Șerban & Todericiu, 2020; Terzopoulos & Satratzemi, 2020). In particular, it has been suggested that AI speakers could be used to tailor learner-centred classroom environments when context-specific features are developed and embedded (Kim, 2018). However, there are few studies that provide a comprehensive perspective on AISEd.

AISEd provides an opportunity to realise the promise of AIEd, using speaker hardware as a mediator between students and AI models. In addition, AI speakers with personalised voices could reduce the psychological distance between human users and AI services (Kim et al., 2018; Kim et al., 2021; Terzopoulos & Satratzemi, 2020). Previous research on AISEd has focused on the potential for native language or English education (Chen et al., 2021; Chung & Woo, 2020; Chang, 2019), as speech-to-text, text-to-speech, and natural language processing technologies enable real-time conversations between students and speakers, which can be effective for language learning. For example, Kim et al. (2021) used a smart speaker on a NUGU platform to improve the metamemory ability of elderly adults as a medical treatment. Renzella et al. (2022) developed the Deep Speaker algorithm to verify the identity of students and integrate speakers with the Real Talk student-tutor discussion and assessment system.

In particular, Authors (2023b) developed a hands-free AI speaker system to support hands-on science laboratory instruction with procedural knowledge - calculating the amount of a reagent, guiding experimental procedures, and instructing where to properly dispose of liquid waste. With the system, students could receive timely information without violating lab safety guidelines (e.g., not touching smart devices or laptops with gloved hands) while concentrating on essential hands-on manipulations. AI speakers can assist students in two ways. First, AI speakers facilitate auditory learning, which has been overlooked in laboratory instruction. Second, students can benefit from the explainable behaviour of AI speakers, as the reason why and how the AI speaker responds to a particular user articulation was structured by the developer, which can mitigate ethical concerns in AIEd (see Hemmer et al., 2021). In particular, the latter is thought to distinguish AI speakers from other AI services (Authors, 2023b).

However, although AI speakers could provide content knowledge with natural language like human teachers/tutors, little research has been conducted on integrating AI speakers to improve the cognitive learning process, especially for science education. Consequently, exploring the possibility of AISEd for cognitive learning in science education can be a touchstone that examines how much a widespread AI service could be transferred to education, foreseeing knowledge sharing in future science classrooms. In our case, we trained the CLAIS system that smart AI speakers can be authentic and heterogeneous peers in the CL context, engaging in linguistic conversation with human peers for knowledge co-construction (Figure 1). The details of the CLAIS system are presented in the Method section.

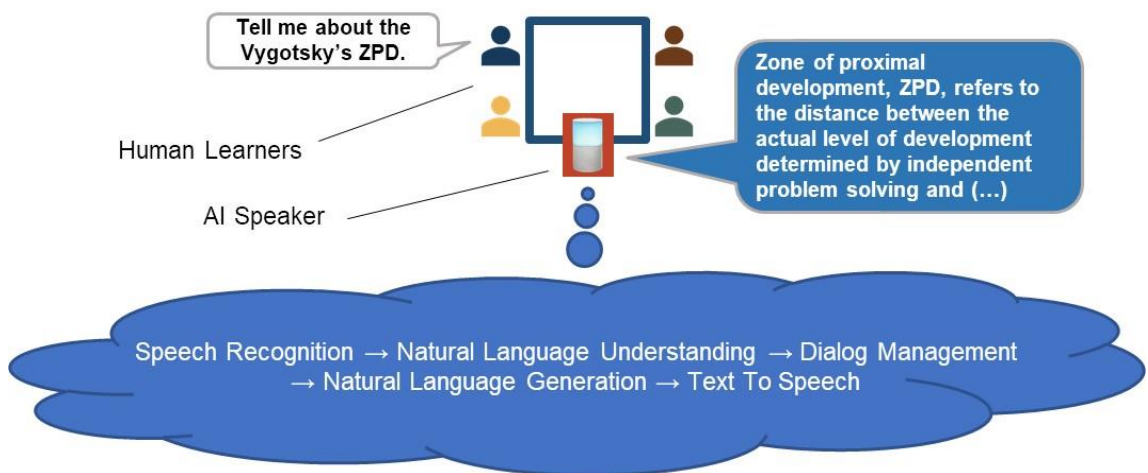

Figure 1. Scheme of the CLAIS system.

*1.5 The Need for Manifold Exploration and Research Questions*

This study responds to the need for research that investigates how epistemic processes can be changed by the introduction of AI in the classroom. Therefore, it is necessary to confirm that AI-integrated authentic teaching and learning is possible with the CLAIS system. This should be followed by a study of students' reactions to their experiences. Since the implementation of AIEd in the classroom in principle affects teachers, students and other stakeholders (Luckin et al., 2016; Hwang et al., 2020; Srinivasan, 2022), it is necessary to investigate the various aspects of the CLAIS system. In this sense, pre-service teachers are appropriate participants as they can report on their experiences as teacher, learner, peer and user. By teacher, we mean the usual K-12 teacher who teaches 20-30 students in a regular classroom according to the given curriculum; by learner, the student within any K-16 educational level who studies a given content in a regular classroom; by peer, the epistemic agent who participates in the co-construction of knowledge in the CL situation; and by user, the person who experiences the affordances and limitations of a system. This study answers five research questions (RQs):

RQ 1. Can the CLAIS system prototype developed in this study be implemented in an authentic teaching and learning situation?

RQ 2. What are pre-service elementary science teachers' perceptions and responses to the CLAIS system prototype as teacher?

RQ 3. What are pre-service elementary science teachers' perceptions and responses to the CLAIS system prototype as learner?

RQ 4. What are pre-service elementary science teachers' perceptions and responses to the CLAIS system prototype as peer?

RQ 5. What are pre-service elementary science teachers' perceptions and responses to the CLAIS system prototype as user?

## 2. Method

### *2.1 Design and Development Research*

D&D research is "the systematic study of design, development and evaluation processes with the aim of establishing an empirical basis for the creation of instructional ... tools" in the field of educational technology (Richey & Klein, 2007). Richey & Klein (2007) categorise D&D research as - Type I: Product & Tool research and Type II: Model research. The former is relevant to this study as it aims to produce a CL system using an AI speaker. In a Type I research, the researcher focuses on the specific product and derives context-specific lessons from its development and analysis of the conditions for its optimal use (Richey et al., 2004). A product development study is typically descriptive research using the case study approach with a mixed-method that collects both quantitative and qualitative data (Richey & Klein, 2007; Richey et al., 2004). D&D research can be a way to establish new tools to support student learning with empirical data (Richey & Klein, 2007) - so applying the D&D approach to the AI speaker system for CL would provide unprecedented insights into its promises and pitfalls.

This study is the initial report of a long-term project to develop and implement the CLAIS system for use in authentic science teaching and learning environments. We have developed a prototype of the CLAIS system and report on pre-service primary science teachers' responses to it, in order to provide timely implications for the field of science education research.

## 2.2 ADDIE Process

D&D studies typically involve the generic Analysis, Design, Development, Implementation and Evaluation (ADDIE) process (Richey & Klein, 2007), which this study followed.

### 2.2.1 Analysis

The need to respond to the call for papers and the literature reviewed above form part of the analysis phase. In addition, the research area, the participants and the available AI speaker platform were analysed. This study took place in a science education course at a national university in Korea. The university is located in Incheon, a metropolitan city in the pan-capital area. The university is dedicated to preparing of pre-service elementary teachers, providing them with an intensive 4-year education to develop teaching knowledge and skills, with curriculums that include educational theories, various subject-matter education and student counselling. Students accepted to these teacher education colleges are usually considered to have higher academic performance compared to their peers. The sophomores who were the participants in this study took basic courses in educational psychology and curriculum and instruction when they were freshmen.

The objective of the course was set as "pre-service teachers learn how to do a scientific inquiry by reading the given materials and practicing scientific writing heuristic template with a primary science activity." The topics covered in the course included the meaning of teaching and learning (week 1), textbook evaluations (week 2), the meaning of science education in Korea (week 3), the nature of science (week 4), the Korean science curriculum (week 5), science process skills (week 6), various teaching models (week 7), learning and development (week 10) and microteaching of pre-service teachers (weeks 9–15). The full course syllabus is summarised in Appendix 1. In weeks 1, 3, 5 and 10, students were introduced to different learning theories such as Piaget's cognitive constructivism, Vygotsky's social constructivism, Ausubel's meaningful learning theory, Bruner's discovery learning theory and Driver's theory of alternative conceptions. A total of 15 sophomore students enrolled in the course and participated in the study. One student only responded to the post-test.

Given the research field and participants in Korea, the potential AI speaker platform had to be able to handle Korean natural language processing. The NUGU platform, which is serviced by SK Telecom in South Korea, was analysed to be used for authentic learning environments with a Korean speaking environment (e.g. Kim et al., 2021; Authors, 2023b).[2] The platform includes a NUGU Play Builder, which allows developers to create custom programs with a natural language understanding (NLU) engine that can be loaded onto an AI speaker. The platform uses the terms *entity* to refer to developer-defined categories of word tokens, *intent* to refer to the user's intention of articulation as inferred by a combination of entities, and *action* to refer to the AI speaker's response that corresponds to each intention.

*2.2.2 Design*

Based on the above analyses, the researchers designed the CLAIS learning system, which incorporates CL and AI speakers. The CLAIS system was based on the Jigsaw CL model, reflecting its strengths above analysed in supporting cognitive learning in higher education, computer-mediated CL, and its highly organised structure that could facilitate the use of the AI component. Also, the AI speaker as one of the most used AI services, the personalised assistant, and the possible 'explainable' teaching tool were considered significant. The overall flowchart of the CLAIS system is shown in Figure 2.

---

[2] https://www.nugu.co.kr/ (Retrieved November 19th, 2023)

- Stage 1 - Home Group: In the Jigsaw teaching and learning model, each 4-6 learners form a home group and individuals take responsibility for the segmented learning content. The teacher explains the procedure of Jigsaw CL.
- Stage 2 - Expert Group: Learners from the different home groups meet as expert groups according to the learning content assigned to individuals. In each expert group, the learners study the same content, as they are designated as experts on the subject for the home group.
- Stage 3 - Home Group: After the expert group learning, the home groups gather again. Each learner in the home group becomes a tutor and explains one by one what they have learned in the expert group.
- Stage 4 - Problem Solving: After all the explanations have been given to each other, the group should have learned all the content. A final assessment with problem solving helps learners to check their learning.

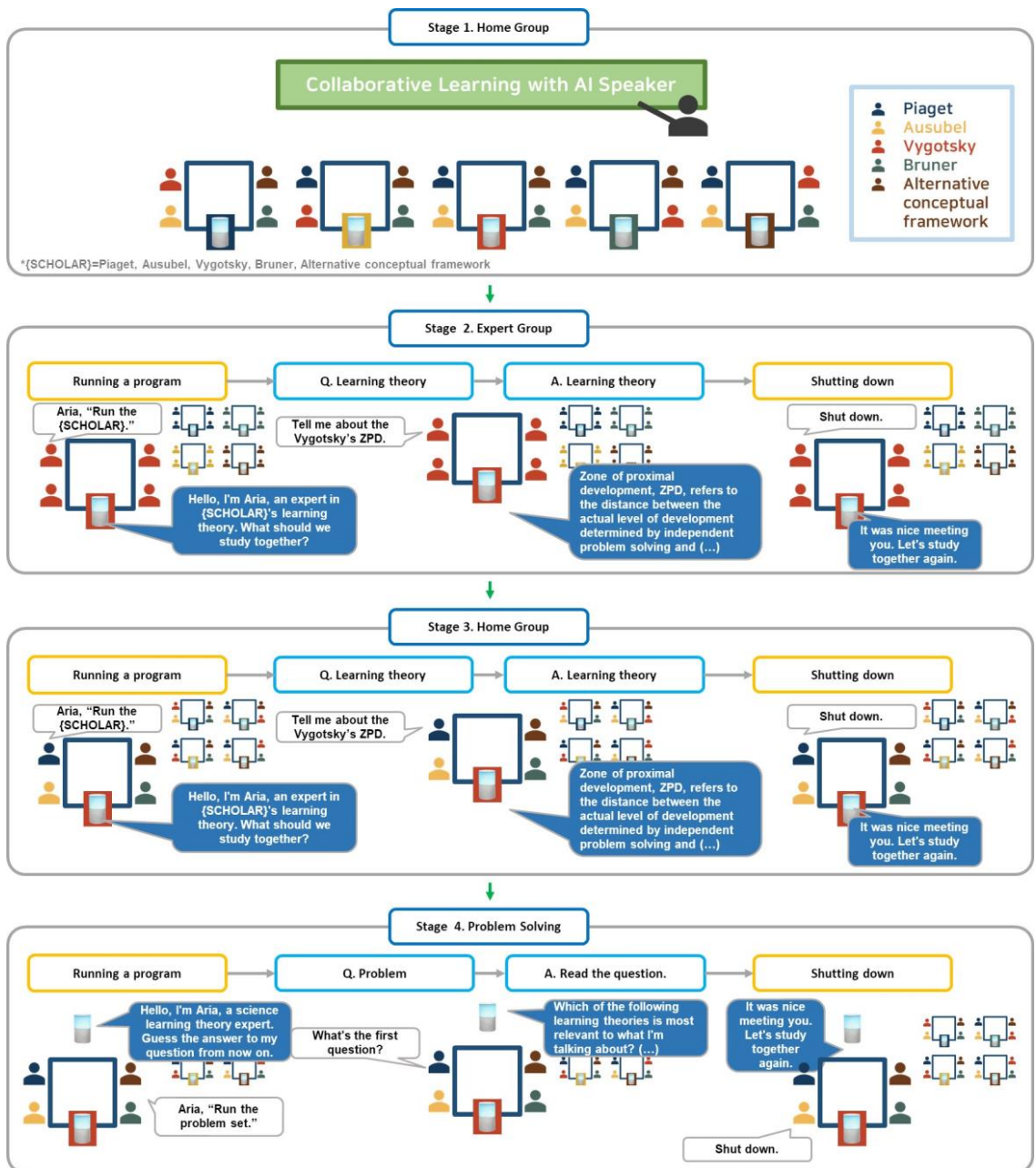

Figure 2. Flow chart of the CLAIS-Jigsaw system

Previously, the Jigsaw CL participants were supposed to be human learners. The novelty of this study is that an AI speaker can be a member of each home group and expert group. Although the teacher continues with Jigsaw CL as before, the students experience AI speakers explaining the part of the Jigsaw process that they have taken over.

The learning content for the prototype of the CLAIS system was a recap of the learning theories of Piaget, Vygotsky, Ausubel, Bruner and Driver. The fill-in-the-blank style was proposed to structure the learning process and simplify the learning task, taking into account the possible difficulty of the low speech recognition performance of the AI speakers in the authentic classroom environment with a lot of noise. Additionally, the learning content-related problem set was considered as a wrap-up.

The design phase of the CLAIS system was guided by three experts in science education who provided expert reviews. Their opinions included that the CLAIS system will provide pre-service teachers with an experience of AI as a collaborator and lead them to reflect on the role of future school teachers. However, they felt that the current state of the CLAIS system would limit its use for simple cognitive learning rather than higher-order thinking or complex interactions.

*2.2.3 Development*

Researchers developed the prototype of the CLAIS system on the NUGU AI speaker platform, as it was thought to be able to satisfy the considerations from the analysis and design phases (e.g. Korean spoken language processing). We defined the required *entities*, *intents* and *actions* in the NUGU Player Builder (Figure 3). The examples of *entities*, *intents* and *actions* for each function are shown in Figure 4. And we trained the NLU model in the NUGU platform by inputting several word tokens for an *entity* and dozens of sentences containing a combination of *entities* for an *intent* (cf. Authors, 2023b).

Figure 3. The CLAIS system developed in the NUGU Play Builder

| Function | Intent (user articulation) | Entity (word tokens) | Action (AI speaker articulation) |
|---|---|---|---|
| (1) Learning | - LEARNING_PIAGET_CALL: "Tell me about the **Piaget no. 1**." → | - PIAGET_ONE: Piaget's theory of cognitive development, Piaget number one, Piaget's first theory → | - ANSWER_PIAGET_ONE: "In **Piaget's theory of cognitive development**, intelligence is a logical thinking process and (…)" |
| | - LEARNING_AUSBEL_CALL: "Tell me about the **Ausbel's third theory**." → | - AUSBEL_THREE: Ausbel's conditions of meaningful learning, Ausbel number three, Ausbel' third theory → | - ANSWER_AUSBEL_THREE: Ausbel suggests learning tasks, cognitive structures, and willingness to learn as conditions for **meaningful learning** where meaningful acceptance learning will occur most effectively. (…) |
| | - LEARNING_Vigotsky_CALL: "Tell me about the **Vigotsky's ZPD**." → | - VIGOTSKY_TWO: Vigotsky's zone of proximal development, Vigotsky;s ZPD, Vigotsky number two, Vigotsky's second theory → | - ANSWER_VIGOTSKY_TWO: **Zone of proximal development, ZPD**, refers to the distance between the actual level of development determined by independent problem solving and (…) |
| | - LEARNING_BRUNER_CALL: "Tell me about the **Bruner's structure of knowledge**." → | - BRUNER_TWO: Bruner's structure of knowledge, Bruner number two, Bruner's second theory → | - ANSWER_BRUNER_TWO: "In Bruner's learning theory, **the structure of knowledge** is described in three ways. First, there are three modes of representations, enactive, iconic, and symbolic. (…)" |
| | - LEARNING_ALTERNATIVE_CALL: "Tell me about the **Subject-specific Alternative conceptual framework**." → | - ALTERNATIVE _THREE: Alternative conceptual framework, Alternative conceptual framewor number three, Alternative conceptual framewor third theory → | - ANSWER_ ALTERNATIVE _THREE: First of all, the field of physics consists of abstract concepts and theories, so many **misunderstandings** have been confirmed, and misunderstandings of power and motion are most comprehensively studied. (…) |
| (2) Assessment | - PROBLEM_CALL: "What's the **first question**? → | - FIRST_QUESTION: Question no.1, First question. → | - ANSWER_PROBLEM_ONE: "Which of the following learning theories is most relevant to what I'm talking about? (…)" |

Arrows(→): Signifies that the recognized word token from the user articulation is transferred to the AI speaker articulation.

Figure 4. Examples of intents, entities, and actions in the CLAIS system

The researchers tested the system in two rounds. (1) The chatbot-like test system in the NUGU Play Builder was used to ensure that the defined functions worked well when the appropriate text was entered. (2) The developed system was loaded onto a NUGU speaker and the researchers made an articulation to invoke the functions. If the system did not respond appropriately, the researchers retrained the Play with additional data.

The researchers also developed a handout containing learning content and problem sets (Figure 5; see Appendix 2 for the English version). The learning content was prepared by summarising the textbook for the *Science Education* course on different learning theories and making blanks. The problem set was prepared by excerpting actual problems from the Korean Teachers' Employment Examination.

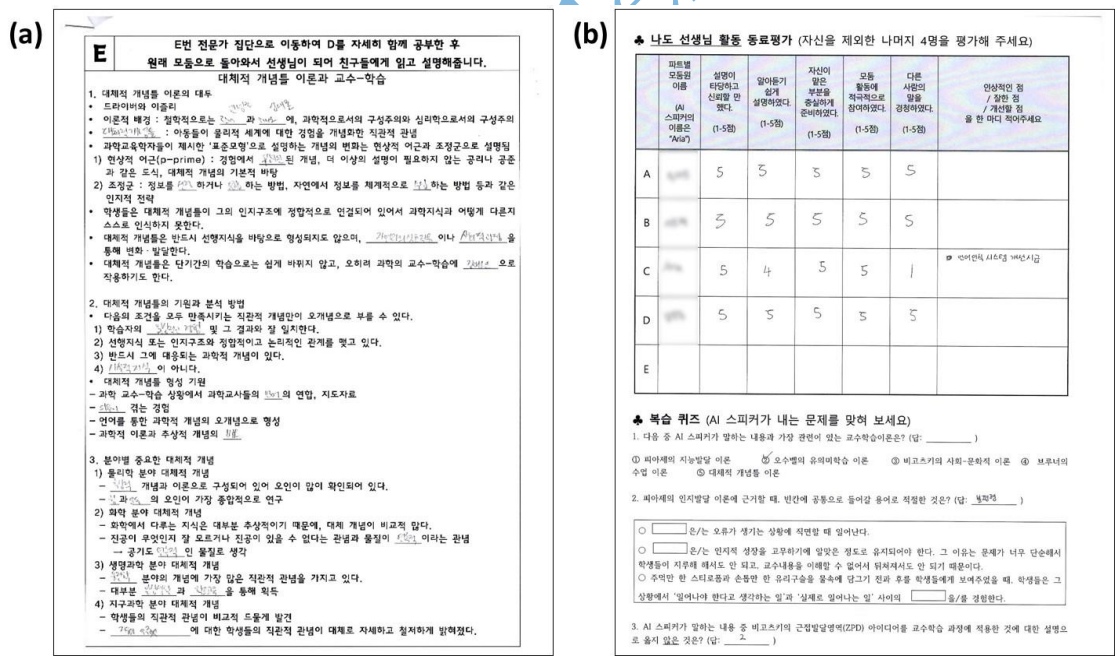

Figure 5. Handout with learning content and problem set (a: a handout page on Driver's alternative conceptual framework, b: peer evaluation rubric and problem set)

*2.2.4 Implementation and Evaluation*

The implementation of the prototype of the CLAIS system took place in the 15th week of the *Science Education* course as a class review and reflection (Appendix 1). The evaluation of the system prototype was carried out using Google Forms with the instruments described below. Students were given about 20 minutes to respond to each survey. The details of implementation and evaluation will be presented in 3. Results and 4. Discussion sections.

*2.4 Instrument*

*2.4.1 Intelligent-Technological, Pedagogical, And Content Knowledge (I-TPACK)*

The items to measure the technological, pedagogical and content knowledge (TPACK) of pre-service primary science teachers considering AI technology were adopted from Celik (2023). He conceptualised the Intelligent-TPACK scale to have 5 dimensions - Intelligent technological knowledge (TK), Intelligent technological pedagogical knowledge (TPK), Intelligent technological content knowledge (TCK), Intelligent TPACK, and Ethics. He developed 27 items corresponding to the dimensions on a 7-point Likert scale and statistically validated the instrument. Participants in this study were also asked about their views on the changes in the role of future teachers due to the development of AI technologies in an open-ended question. The I-TPACK questionnaire was administered twice - before and after the implementation of the CLAIS system.

*2.4.2 Perceptions of an AI Speaker System in Collaborative Learning (PASS-CL)*

The items to investigate participants' perception of the AI speaker system as learner were adopted from Authors (2023b). They designed the Perception of an AI Speaker System in Science Laboratory Classroom (PASS-SLC) questionnaire to investigate students' perception of the hands-free AI speaker system supporting hands-on science laboratory classes, based on the literature on technology acceptance model and usability of AI speakers or chatbots (Authors, 2023b). The questionnaire consists of 18 items on a 5-point Likert scale. The researchers made slight changes to the wording of the items (changing "laboratory class" to "collaborative learning") to match the research context of this study. In addition, the open-ended question asked participants about their views on the changes in the role of future learners due to the development of AI technologies.

*2.4.3 Peer Evaluation for Collaborative Learning*

The items for the peer evaluation of human and AI peers were adopted from the literature that aimed to improve the quality of CL through peer assessment of the learning process (Divaharan & Atputhasamy, 2002; Hu et al., 2022; Jeong & Park, 2022; Choi, 2022). The items asked participants whether a peer "provided valid and reliable explanation," "explained in an easy-to-understand way," "prepared the part he/she took responsibility of," "actively participated in the group activity," and "listened carefully to what others say" on a 5-point Likert scale. Participants were also asked to comment on a peer if there were any impressive/good/improvement points.

*2.4.4 People At the Centre of Mobile Application Development (PACMAD)*

The items for the usability test were adopted from Harrison et al. (2013). Their PACMAD (People At the Centre of Mobile Application Development) usability model consisted of three factors, user, task and context of use, which included seven attributes of usability: effectiveness, efficiency, satisfaction, learnability, memorability, error and cognitive load. Although originally designed for mobile applications, the PACMAD model has been used to test the usability of mobile loudspeaker systems (Shaikh, 2020; Xiao et al., 2020; Authors, 2023b). The researchers asked participants to rate each of these seven attributes of the AI speaker system for CL on a 1-4 point Likert scale, and to explain why they did so for each attribute. In the open-ended questions, the researchers also asked them about the pros and cons of the system and suggestions for revising the system.

*2.5 Data Analysis*

*2.5.1 Quantitative Data*

Descriptive statistics were obtained for participants' responses. For inferential statistics, non-parametric statistics had to be used due to the small sample size ($N = 15$). Therefore, the Wilcoxon matched-pairs signed-rank test and the Wilcoxon rank-sum test (Mann-Whitney two-sample statistic) were used for statistical inference. STATA 16 was used throughout the quantitative data analysis.

*2.5.2 Qualitative Data*

The qualitative data came from the open-ended questions corresponding to I-TPACK, PASS-CL and PACMAD. The qualitative data were analysed by three coders - an expert and a Master's student in chemistry education and an expert in earth science education. As the qualitative data was intended to elaborate on the quantitative responses of the participants, the coders relied on the relevant questions in the survey. However, they extracted lower-level codes to explain the patterns in the quantitative data.

The videotaped sessions of the implementation of the AI speaker system were transcribed. The researchers observed the data, focusing on the verbal interaction between the students and the AI speaker system. The videos and transcripts were used to triangulate our data analysis.

## 3. Results

### *3.1 The Implementation of the CLAIS System Prototype*

The developed CLAIS system prototype was successfully implemented in the last session of the *Science Education* course, where 15 students participated for 60 minutes. A total of five groups were formed, each consisting of an AI speaker and 3-4 pre-service elementary science teachers. The course instructor first introduced the CLAIS system and explained how to proceed with learning within it. The students easily followed the instruction to proceed with CL while an AI speaker was considered as a member of each group. Figure 6-(a) shows the view of the classroom where the CLAIS system was implemented, while Figure 6-(b) shows the learning in the expert group and Figure 6-(c) the learning in the home group.

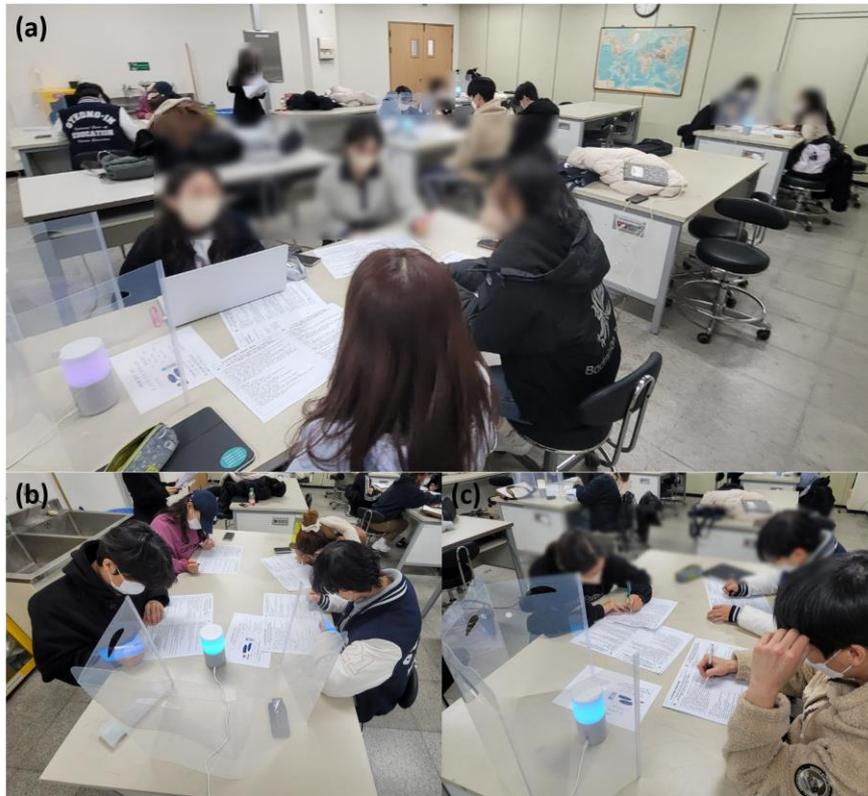

Figure 6. The implementation of the CLAIS system prototype (a: overall classroom, b: expert group learning, c: home group learning)

Throughout the CLAIS-Jigsaw teaching and learning process, students in a group could interact with peers, including human students and an AI speaker, asking questions about the learning content (learning theories of scholars) and listening to the peer responsible for specific content respond. The example dialogue is shown below (note that the system call name is 'Aria', which is provided by the platform):

Student: "Aria? (system rings) Tell us point number two of Piaget's learning theory."

> Aria: "The first stage of intelligence development corresponds to children 0–2 years old. In this period, the sensorimotor schema develops, and verbal and cognitive schemas are small and highly unstable. …"

However, sometimes the AI speakers did not recognise the articulation of the human peer because the surrounding noise made the user's voice indistinguishable. The speaker would then ask the student to repeat the question. If this happened repeatedly, the teacher intervened to ask the AI speaker instead of the student, or to restart the system. The example dialogue is shown below:

> Student: "Aria? (system rings) Tell us point number three (of Vygotsky's learning theory)"
> Aria: "Would you please say it again?"
> (the same conversation repeats five times)
> Student: "Point number three!"
> Aria: "In Vygotsky's theory, the need and methods of dynamic science assessment are suggested. According to Vygotsky, a language largely influences mental development, such as the development of scientific conception …"

In the final stage, an instructor asked an AI speaker to present problems to the students, and the students listened to what the AI speaker said and answered the problems in the handouts. The dialogue is shown below:

> Instructor: Aria. (system rings) Give us question number one.
> Aria: Which of the following learning theories is most relevant to what I'm talking about? It emphasises the incorporation of new knowledge or information with prior knowledge, … (students write down their answers)

### *3.2 Pre-service Elementary Science Teachers' Response as Teacher*

Table 1 shows the results of the pre- and post-tests of I-TPACK. Wilcoxon matched-pairs signed rank test was conducted to compare the pre- and post-tests on I-TPACK for both the overall scores and the scores on individual dimensions. Results indicate that the post-test on overall I-TPACK (M = 4.97, SD = 1.05) was significantly higher than the pre-test (M = 3.43, SD = 1.26) $Z$ = 3.296, $p < .01$. This result suggests that the implementation of CLAIS did significantly increased participating teachers' I-TPACK. Wilcoxon matched-pairs signed rank test for I-TPACK also found statistical differences between post- and pre-tests scores for each dimension, at the $p < .01$ level.

Table 1. The I-TPACK survey result (1-7 scale) ($N$ = 14)

| Category | Pre-test | Post-test | $Z$ |
| --- | --- | --- | --- |
| AI_TK | 3.86 (1.45) | 4.87 (1.06) | 2.891 ** |
| AI_TPK | 3.77 (1.51) | 4.95 (1.04) | 2.954 ** |
| AI_TCK | 3 (1.47) | 4.79 (1.07) | 3.028 ** |
| AI_TPACK | 3.05 (1.27) | 5.04 (1.26) | 3.301 ** |
| AI_Ethics | 3.53 (1.32) | 5.16 (1.05) | 3.246 ** |
| **Average** | **3.43 (1.26)** | **4.97 (1.05)** | **3.296** ** |

** $p < .01$

The pre-service teachers' open-ended responses about their expectations for the role of future teachers anticipated substantial changes in it - some insisted that teachers should be able to control or collaborate with the AI-embedded system, while others expected teachers to do more humane work, such as emotional care. And their answers tended to be elaborated after experiencing the CLAIS session. Representative quotes are as follows,

(Before CLAIS) "Teacher as content deliver would diminish."
(After CLAIS) "The role of knowledge deliverer would diminish. The role of giving feedback and overall management and implementation of learning facilities will be strengthened."

(Before CLAIS) "Teachers should raise competency to utilise AI in instruction, and their role will change to helping"
(After CLAIS) "Teachers will have abilities to utilise AI appropriately and will take the role of instructor together with AI."

(Before CLAIS) "[no response]"

(After CLAIS) "[Teacher will be] a person who communicates emotions with children."

### *3.3. Pre-service Elementary Science Teachers' Response as Learner*

The result of the PASS-CL questionnaire survey after the implementation of the CLAIS system is presented in Table 2. Although the overall score of 4.06 (SD = 0.55) was higher than 4, there were differences between the scores of the individual items. The lowest score came from item 2 (2.87, SD = 1.19), the only item to score below 3. This is related to the speech recognition performance of the AI speakers and will be explained later. Meanwhile, the scores for perceived accuracy (no. 3), consistency (no. 4) and reliability (no. 5) of the AI speaker were remarkably high, being close to or above 4.5. Students' perceptions of the changed learning process and interactions (no. 13-15) were also high with scores above 4. Furthermore, students' perceived interest (no. 17) and enjoyment (no. 18) also scored more than 4. This shows that participants as learners were generally satisfied with the CLAIS system and in particular perceived it as correct, consistent, reliable, fun and enjoyable. They also felt that it changed the learning process in the classroom.

Table 2. The PASS-CL survey result (1-5 scale) ($N = 15$)

| Item no. | Question | Score (mean [sd]) |
| --- | --- | --- |
| 1 | I was able to ask the speaker in the way I usually speak. | 3.8 (.94) |
| 2 | The speaker understood the intention of my question well. | 2.87 (1.19) |
| 3 | The answer of the speaker was easy to understand. | 3.93 (.88) |
| 4 | The speaker told the correct information needed for the class. | 4.47 (.64) |
| 5 | The information the speaker told was consistent. | 4.6 (.63) |
| 6 | The information the speaker told was reliable. | 4.67 (.62) |
| 7 | The system was more convenient for getting the information I wanted than using calculators or smartphones. | 3.33 (.18) |
| 8 | The class became safer than before through this system. | 4.07 (.8) |
| 9 | Using the system was more comfortable than asking friends or TA. | 3.73 (1.22) |
| 10 | As a result of using the system, I could gain more scientific knowledge. | 3.87 (.92) |
| 11 | As a result of using the system, I could gain more scientific skills. | 3.93 (.59) |
| 12 | As a result of using the system, I could gain more attitude toward science class. | 3.93 (.88) |
| 13 | The process of science classes using the system became different from before. | 4.33 (.9) |
| 14 | The role of the instructor has changed in the science class using the system. | 4.13 (.99) |
| 15 | The role of the students has changed in the science class using the system. | 4.2 (.77) |
| 16 | The learning activity time has been reduced when using the system. | 4.29 (.72) |
| 17 | The user experience of the system was fun. | 4.53 (.74) |
| 18 | The user experience of the system was enjoyable. | 4.53 (.52) |
| | Overall | 4.06 (.55) |

In the open-ended questions about interactions during the CLAIS system, students responded that AI had become the mediator of knowledge delivery, while at the same time bringing about changes in the roles of teacher and student, and thus the aspects of teacher-student and student-student interactions in the classroom. Representative quotes are as follows:

> "Before the use [of the system], the conversation between instructor and students were more direct. While using this system, instructor was changed to explain the method to use the system for learning. The conversation between students was consist of how to use the system effectively."
>
> "The class was proceeded in a teacher-AI-student way."
>
> "The instructor did not provide many explanations but changed to help only when there are errors in the system."
>
> "The instructor does not deliver all the knowledge and students deliver with each other what they've got to know through AI."
>
> "The speaker became a model for explaining - so interaction between students as expert was smoother."

### *3.4 Pre-service Elementary Science Teachers' Response as Peer*

The result of the peer assessment on the human and AI peers is presented in Table 3. It should be noted that the response patterns of the pre-service elementary science teachers on the peer evaluation sheet differed between human and AI peers. Most students gave 5 points to their human peers in the collaborative group (M = 4.98; SD = .09). Participants generally rated the AI peer positively (M = 4.34; SD = .89 on a 5-point Likert scale) - they even rated the AI speaker's response to their questions as honest preparation for the part it was responsible for (M = 4.64; SD = .93) and active participation in the Jigsaw activity (M = 4.35; SD = 1.01). Wilcoxon matched-pairs signed rank test was conducted to compare the students' peer assessment on AI speaker and human peers. Results indicate that the scores of the AI peers were significantly lower than those of the human peers in the overall sense ($Z = 4.742$, $p < .001$). Wilcoxon matched-pairs signed rank test for the peer assessment also found statistical differences between the AI speaker and human peers for each dimension, at the $p < .05$ or less level. Note that the lowest score was for whether the AI speaker listened carefully to what others were saying (M = 3.71; SD = 1.44), which was less than 4.

Table 3. The result of peer assessment on AI speaker and human peers

| Question | Overall ($N = 54$) | On AI speaker ($n = 14$) | On human ($n = 40$) | Z |
| --- | --- | --- | --- | --- |
| Validity and reliability | 4.91 (.49) | 4.64 (.93) | 5 (0) | 2.413 * |
| Easy to understand | 4.83 (.54) | 4.36 (.93) | 5 (0) | 4.344 *** |
| Prepared the part he/she took charge of honestly | 4.91 (.49) | 4.64 (.93) | 5 (0) | 2.413 * |
| Actively participated in group work. | 4.76 (.67) | 4.35 (1.01) | 4.9 (.44) | 2.876 ** |
| Listened carefully to others | 4.67 (.91) | 3.71 (1.44) | 5 (0) | 5.112 *** |
| Overall | 4.81 (.53) | 4.34 (.89) | 4.98 (.09) | 4.742 *** |

\* $p < .05$, \*\* $p < .01$, \*\*\* $p < .001$

The difference in the response pattern for peer evaluation of AI and human is also evident in the non-compulsory comment on the peer evaluation form. For human peers, there were 5 comments out of 40 observations (12.5%). And their points were simply complimentary of the human peer, mentioning 'he/she is smart', 'he/she explained well',

'the interaction was smooth' and 'the feedback was immediate'. In contrast, for the AI peer, there were 10 comments out of 14 observations (71.4%). Of these, seven comments mentioned that the AI speaker did not do a good job of recognising what the students were articulating. Two other comments mentioned that the AI speaker was slow to respond to their query. Another comment said that he/she had to be in tension to listen to what the AI speaker was saying. Representative quotes are as follows:

> "It does not well understand what I say. It was difficult to recognise words when I am listening to unfamiliar part."
> "Its response is still slow."
> "It was impossible to stop it saying - so I had a tension as much as listening assessment."

### *3.5 Pre-service Elementary Science Teachers' Response as User*

The result of the usability test based on the PACMAD questionnaire is shown in Table 4. The overall mean score of 3.14 (SD = .45) was positive (> 3) on a scale of 1-4. However, 'satisfaction' in the 'task' category (M = 2.93, SD = .88) and 'cognitive load' in the 'content' category (M = 2.53, SD = 1.36) scored relatively low (< 3) compared to the others.

Table 4. The PACMAD usability survey result (1-4 scale) ($N = 15$)

| Category | Item | Score (mean [SD]) |
|---|---|---|
| User | Effectiveness | 3.27 (.59) |
| | Efficiency | 3.2 (.68) |
| | Average | 3.23 (.59) |
| Task | Satisfaction | 2.93 (.88) |
| | Learnability | 3.6 (.63) |
| | Average | 3.27 (.65) |
| Content | Memorability | 3.47 (.92) |
| | Error | 3 (.93) |
| | Cognitive load | 2.53 (1.36) |
| | Average | 3 (.51) |
| Overall average | | 3.14 (.45) |

Participants explained their answers in open-ended questions corresponding to the PACMAD categories. For the 'user' category, most responses were positive:

"The fast and accurate answer of the AI"

"The information is quite accurate and easy to accept."

However, for the 'task' category, respondents specified why they had not been satisfied with the system much:

"The device sometimes did not understand the learner's saying. However, it was enjoyable to learn how to use it."

For the 'content' category, respondents were ambivalent about good memorability but possible cognitive load:

"How to use it was simple, and if the user's [operational] content is successfully delivered, it will be available without error. However, I had a lot to think about what to ask for learning."

"It is easy to memorise how to use it. However, while listening to what the AI speaker says, non-verbal communicative factors are not reflected. And the speed of [speaker's] articulation was fast, causing the cognitive load."

Finally, respondents made suggestions for a future revision of the CLAIS system. These included "encouraging mutual interaction", "giving choices", "adapting to students' level", "gamification" and "recoding [students'] questions".

Note that the data reported in the 3. Results section is intended to be considered as part of the Implementation and Evaluation stages of the ADDIE process, which is also the case for 4. Discussion.

## 4. Discussion

## 4.1 RQ 1: The Characteristics of the Implemented CLAIS System Prototype

The CLAIS system prototype developed in this study could be implemented in the 60-minute *Science Education* lesson. The human learners and the AI speaker became equivalent members of the group and took part in the Jigsaw CL process - i.e. each became a member of the home and expert groups explaining the divided learning content to the others. After each member has finished their explanation, each learner in the group has to complete their learning tasks.

This study showed that human learners and AI could work together as a group for knowledge co-construction in authentic teaching and learning situations. It clearly shows that knowledge is an artifact that is constructed through collaboration not only in a group of humans but also in a group of humans and AI. Therefore, the successful implementation of the CLAIS system prototype has shown how developing AI technologies are changing the shape of classrooms, which is not a distant future.

## 4.2 RQ 2-5: Pre-service elementary science teachers' responses to the CLAIS

In this study, the participating pre-service teachers evaluated their multifaceted experiences with the CLAIS system prototype. First, we found that the CLAIS system could improve the participating pre-service teachers' TPACK in terms of AI. This finding was evidenced by the respondents' significantly increased scores on I-TPACK after the 60-minute implementation of the system (Table 1). This seems to be because the pre-service elementary science teachers were not familiar with the AI-based teaching and learning platform, and the CLAIS system provided them the opportunity to experience with AI. Therefore, our findings emphasised that the CLAIS system can be a candidate to promote the I-TPACK of pre-service or in-service science teachers who have limited experience using AIEd in their classrooms. In particular, based on the participants' responses to the open-ended questions, we content that the CLAIS system can be used to invite teachers to reflect on the "roles" of teachers in the era of AI. If AI could serve as a heterogeneous and knowledgeable peer extending ZPD in a small group, teachers may find their new roles to design and manage such a learning environment using AI in the classroom.

Second, the respondents as learners showed that they perceived the AI speaker as giving trustworthy information, changing classroom interactions and being enjoyable (Table 2). Indeed, the information given by the AI speaker in the CLAIS system prototype can be considered trustworthy, as the researchers had used information extracted from the Science Education textbook to train the AI speakers. However, students must also be aware of the possible bias in the AIEd models and become critical evaluators of the information (Kim, 2022; Lai et al., 2022). Therefore, in the future CLAIS system or other CL systems with AI, consideration should be given to fostering students' critical evaluation skills. Furthermore, although this study focused on the perceived changes in learner interactions, more qualitative studies focusing on the dialogue between human learners and AI should follow to answer in depth the question of how AI can change the collaborative process when considered as a member of the group. The fact that the CLAIS system has aroused learners' interest in science learning is a good reason to continue its development.

Third, respondents as peers showed that human peers and AI peers are different to students (Table 3). We can interpret this phenomenon by adopting the concept of 'meaningful other' - whether or not AI is a 'meaningful other' to students may influence their response to AI in the learning process (Park et al., 2021). It is speculated that respondents in this study may have considered human peers as 'meaningful others' that they are reluctant to evaluate openly, but may have treated AI peers as mere agents that they do not hesitate to evaluate. The question remains as to whether or not students should treat human peers and AI peers equally. Nevertheless, this study has brought the innovation of trying to evaluate human peers and AI peers within the same peer evaluation rubric.

Fourth, the respondents as users showed low satisfaction and high cognitive load while using the CLAIS system prototype (Table 4). Therefore, the directions for the future development of the CLAIS system should aim at increasing user satisfaction and decreasing cognitive load. The main factor that affected user satisfaction and cognitive load was identified as the low speech recognition performance due to noise in the authentic teaching and learning classroom. This issue may be resolved as performance improves with updates to the NUGU AI speaker platform; however, this does not provide meaningful implications for future research. As in a similar previous study (Authors, 2023b), a keyword-based short conversation between human learners and AI speakers might alleviate the performance problem. Similarly, adding more structure to the CL process may also solve the problem. However, in both cases the interaction between human learners and AI speakers may be reduced or simplified. Therefore, " methods for supporting collaboration between humans and AI on shared functions " (NASEM, 2022, p. 44) need to be explored in the future, while improving their CL performance.

### *4.3 Limitations and Future Work*

This study has several limitations. First, because this study focused on the D&D of the prototype of the CLAIS system, we only recruited a small sample of participants ($N = 15$) for a short period of intervention. This design limits the generalizability of the conclusions made on the findings. A follow up study with a larger sample size and a durable intervention period is needed to further examine the potential of the CLAIS prototype to transform the epistemic process in science classrooms. In particular, given that the content delivered in this study was pedagogical knowledge in science education, the scalability of the CLAIS system may allow its future application to science content knowledge.

Our findings also suggested that in future studies it is essential to make the AI speaker 'listen' to the students' articulations. Some technical suggestions include: (1) arranging a quiet environment so that the AI speakers can easily recognise students' speech, (2) preparing error handling for the case when the speakers do not recognise, and (3) constantly updating the synonyms/pronunciations of user articulations observed in the authentic situations to the internal natural language model of the AI speakers. Pedagogically, the CLAIS system can benefit from lower noise when being used for small numbers of students, and teachers could introduce students to the characteristics of AI speakers before using it for instruction. Keyword-based conversation is also a recommended developmental practice (Authors, 2023b).

Some may question what the AI learned from the CL-like session. This is because the researchers of this study aimed to support student learning using a machine, which is the basic approach of CSCL. Apparently, the AI speaker did not undergo cognitive learning during the CLAIS session. However, it can be said that knowledge construction took place within each small group as an activity system. Also, the students' suggestions for the revision of the system can be taken into account while the developers update the program loaded in the AI speaker. In the future, the AI speaker will also learn something during the session in real time, and one of the learning can be about the states of the human learners (Luckin et al., 2016). In the context of authentic teaching and learning, the learner information heard by the AI speaker could be transmitted to a proxy server to be processed and the results returned to the speaker. This will be a topic for future research and development.

Furthermore, the CLAIS session implemented in this study has limited to do with human decision-making, which has drawn concerns about the uses of AIEd (Hwang et al., 2020; Celik, 2023; cf. NASEM, 2022). Research has consistently suggested that teachers' roles could not be substituted by AI (Authors, 2023c). As the human taking charge of the group and responsibility for the decisions is crucial in HAC (NASEM, 2022, p. 13), future teachers practising CL with AI should reflect on the issue of students' responsibility in the group and also invite students to do so. This also calls for another research agenda--future research should analyse the discourse and the dialogue between students and AI. This task will enhance our understanding of the epistemic difference between knowledge constructed only between humans and between humans and AI.

## 5. Conclusion

In this study, the researchers developed and implemented the prototype of the AI speaker-integrated system that enables CL between human learners and AI. The responses of pre-service elementary science teachers as teacher, learner, peer and user were reported and discussed. This initial trial has successfully demonstrated that CL, an important teaching and learning method for knowledge construction in science education, can be transformed with the participation of AI. It is expected that CL with AI will become a common practice in the future, changing the epistemic process of cognitive learning. As researchers and developers, the researchers of this study will continue to develop the system based on the lessons learned from this initial report.

**Appendix 1. Course Syllabus**

| Week | Topics |
|---|---|
| 1 | Orientation for the lecture<br>What teaching and learning means |
| 2 | 1. Work for microteaching preparation<br>2. Textbook evaluations |
| 3 | What science education means in Korea? |
| 4 | - Students as scientists (reading materials)<br>- Nature of science activity |
| 5 | - Korean science curriculum<br>- Presenting your own class activity scheme<br>- Guideline for the presentation and practice<br>- 5E model and its adoption-teaching model |
| 6 | - Science process skills |
| 7 | - Various teaching model |
| 8 | - Process skill 1<br>- Structure of science knowledge |
| 9 | - Process skill 2<br>- Group 1: Presentation of their own designed science activity |
| 10 | - Learning and development<br>- Group 2: Presentation of their own designed science activity |
| 11 | - Group 3: Presentation of their own designed science activity<br>- Group 4: Presentation of their own designed science activity |
| 12 | - Group 5: Presentation of their own designed science activity<br>- Group 6: Presentation of their own designed science activity |
| 13 | - Group 7: Presentation of their own designed activity<br>- Group 8: Presentation of their own designed activity |
| 14 | - Group 9: Presentation of their own designed activity<br>- Group 10: Presentation of their own designed activity |
| 15 | Class review and reflection |

**Appendix 2. English Version of Handout with Learning Content and Problem Set (Figure 5)**

(a): A handout page on Driver's alternative conceptual framework [(**Bold**): The content students have to fill in]

| E | Move to the E expert group and study D in detail. Afterwards, return to your original group and become a teacher to read and explain to your friends. |
|---|---|

### Alternative Framework Theory and Teaching-Learning

1. Introduction to the Alternative Conceptual Framework Theory
- Driver and Easley
- Theoretical background: Philosophically, it's based on (**idealism**) and (**relativism**), and on scientific constructivism and psychological constructivism.
- (**Alternative conceptual framework**): Children's intuitive notions conceptualized from their experiences of the physical world.
- Concepts changing as described by 'standard models' proposed by science education scholars are explained by phenomenal primitives and coordinative classes.
  1) Phenomenal primitives (p-prim): (**Abstracted**) concepts from experiences, schematics like axioms or postulates that don't need further explanation, basic foundations of alternative concepts.
  2) Coordinate clause: Cognitive strategies such as methods to (**select**) or (**integrate**) information, and systematic ways to gather information from nature.
- Students often don't realize how their alternative conceptual frameworks differ from scientific knowledge because these frameworks are coherently connected to their cognitive structures.
- Alternative conceptual frameworks aren't necessarily formed based on prior knowledge; they evolve and develop through (**interaction with nature**) and (**social processes**).
- Alternative conceptual frameworks do not easily change with short-term learning and can even act as (**obstacles**) in the teaching and learning of science.

2. Origins of the Alternative Conceptual Framework and Methods of Analysis
- Only intuitive concepts that meet all of the following conditions can be called misconceptions:
  1) It matches well with learners' (**daily experiences**) and their outcomes.
  2) It has a logical relationship that is consistent with prior knowledge or cognitive structures.
  3) There is a corresponding scientific concept.
  4) It's not (**factual knowledge**).
- Origins of the formation of alternative conceptual frameworks:
  – (**Association**) of language used by science teachers in science teaching-learning situations, and instructional materials.
  – (**Repeated**) experiences.
  – Formation of misconceptions of scientific concepts through language.
  – (**Analogies**) of scientific theories and abstract concepts.

3. Important Alternative Concepts in Various Domains
  1) Physics
    – Consists of (**abstract**) concepts and theories, many misconceptions have been identified
    – Misconceptions about (**force**) and (**motion**) have been most comprehensively studied.
  2) Chemistry
    – Knowledge in chemistry is mostly abstract, hence there are relatively many alternative concepts
    – Notions like not knowing what vacuum is or believing that vacuum cannot exist, or the idea that matter is (**continuous**) → thinking of air as (**continuous**) matter.
  3) Life Sciences
    – Contains the most intuitive concepts in the field of (**genetics**).
    – Most are acquired through (**daily life**) and (**school education**).
  4) Earth Science
    – Intuitive concepts in students are relatively rare
    – Students' intuitive concepts about (**the relationship between the Earth and the universe**) have been thoroughly and precisely identified.

(b): Peer evaluation rubric and problem set

♣ I AM A TEACHER TOO!: PEER EVALUATION (Please evaluate the other 4 members excluding yourself)

| | Name of Member in Each Part (The AI speaker's name is "Aria") | The explanation was valid and trustworthy. (1-5 points) | Explained in an easy-to-understand manner. (1-5 points) | Prepared their assigned part thoroughly. (1-5 points) | Actively participated in group activities. (1-5 points) | Listened attentively to others. (1-5 points) | Please write a brief comment on the following: Impressive point / What was done well / Areas for improvement |
|---|---|---|---|---|---|---|---|
| A | | | | | | | |
| B | | | | | | | |
| C | | | | | | | |
| D | | | | | | | |
| E | | | | | | | |

♣ **Review Quiz** (Guess the answer to the questions posed by the AI speaker)

1. Which of the following learning theories is most related to what the AI speaker says? (Answer: _____)

① Piaget's Theory of Cognitive Development ② Ausubel's Theory of Meaningful Learning ③ Vygotsky's Socio-cultural Theory ④ Bruner's Theory of Instruction ⑤ Alternative Conceptual Framework Theory

2. Based on Piaget's Theory of Cognitive Development, what term appropriately fills in the blanks? (Answer: _____)

○ _____ occurs when faced with situations where errors arise.
○ _____ should be maintained at a level appropriate to stimulate cognitive growth. This is because it's not good if the problem is too simple that students become bored, nor is it good if they fall behind because they can't understand the instructional content.
○ When showing students a foam as big as a fist and a marble-sized glass bead before and after soaking them in water, students experience the difference between what they 'think should happen' and 'what actually happens' in that situation.

3. Among the contents spoken by the AI speaker, which of the following is not an accurate description applying Vygotsky's Zone of Proximal Development (ZPD) idea to the teaching-learning process? (Answer: _____)

**References**

Authors (2019).

Authors (2020).

Authors (2023a).

Authors (2023b).

Authors (2023c).